\documentclass[10pt,conference]{IEEEtran}
\usepackage{amsfonts}
\usepackage{amssymb,epic,eepic}
\usepackage{amsmath}
\newcommand{\cc}{{\mathbb C}}
\newcommand{\rr}{{\mathbb R}}
\newcommand{\mm}{{\mathbb M}}
\newcommand{\idn}{\mathbf{1}}

\newtheorem{theorem}{Theorem}[section]         
\newtheorem{lemma}[theorem]{Lemma}             

\begin{document}

\title{On the Boundedness of the Support of Optimal Input Measures for Rayleigh Fading Channels}

\author{Jochen Sommerfeld$^{1}$, Igor Bjelakovi\'c$^{1,2}$, and Holger Boche$^{1,2}$\\[2mm]
\small $^{1}$Heinrich-Hertz-Chair for Mobile Communications\\ Technische Universit\"at Berlin\\Werner-von-Siemens-Bau (HFT 6), Einsteinufer 25, 10587 Berlin, Germany\\
\&\\
$^{2}$Institut f\"ur Mathematik\\
Technische Universit\"at Berlin\\
Stra\ss e des 17. Juni 136, 10623 Berlin, Germany\\
Email: \{jochen.sommerfeld, igor.bjelakovic, holger.boche\}@mk.tu-berlin.de 
\thanks{This work is supported by the Deutsche Forschungsgemeinschaft DFG via project BO 1734/16-1 ''Entwurf von''.}}
\maketitle
\begin{abstract}
We consider transmission over a wireless multiple antenna communication system operating in a Rayleigh flat fading environment with no channel state information at the receiver and the transmitter with coherence time $T=1$. We show that, subject to the average power constraint, the support of the capacity achieving input distribution is bounded. Moreover, we show by a simple example concerning the identity theorem (or uniqueness theorem) from the complex analysis in several variables that some of the existing results in the field are not rigorous. 
\end{abstract}
\section{Introduction}
We show in this paper by elementary means that the support of the capacity achieving input measure for multiple-input multiple-output (MIMO) Rayleigh fading channels subject to average power constraint with coherence time $T=1$ is bounded. A generalization of the result to coherence intervals of size $T>1$ seems to be highly non-trivial and will probably require a substantial extension of the techniques used here supplemented by some results and methods from the ``hard analysis''. \\
Previous fundamental achievements, e.g. \cite{abou-faycal, chan, fozunbal-a, fozunbal-b}, follow the same procedure which can be traced back to the classic paper \cite{smith} by Smith. The basic tools are the Karush-Kuhn-Tucker (KKT) conditions from the theory of convex optimization supported by an application of the identity theorem (also known as the uniqueness theorem) from complex analysis. Our approach is based on the KKT conditions too but avoids the usage of the identity theorem.\\
In \cite{abou-faycal} Abou-Faycal, Trott, and Shamai proved, using these techniques, that for a one-dimensional Rayleigh fading channel the optimal input measure subjected to an average power constraint to be discrete with a finite number of mass points.  In \cite{chan} Chan, Hranilovic, and Kschischang showed for a MIMO Rayleigh block-fading channel with i.i.d. channel matrix coefficients that the optimum input distribution subjected to peak and average power constraint contains a finite number of mass points with respect to a specific norm. In addition Fozunbal, Mclaughlin, and Schafer argued in \cite{fozunbal-a}  that a bounded support of the capacity maximizer implies its singularity with respect to the Borel-Lebesgue measure. The approach in \cite{fozunbal-a, chan} is based on the identity theorem for holomorphic functions in several complex variables and use the assumption that an open set in $\rr^n$ fulfills the hypothesis of the identity theorem in $\cc^n$. We show  in section \ref{sec-discussion} by a simple example that the conclusion of the identity theorem fails in this setting. Consequently, these results are not rigorous. Since, in contrast to the complex analysis in one variable, it is still an open difficult problem to characterize the families of sets for which the identity theorem for holomorphic functions in several complex variables holds we cannot hope to understand the properties of the capacity maximizers in the present setting by an reduction to uniqueness properties of holomorphic functions in higher dimensions. Therefore, it is likely that we will be forced to develop or apply ``real-analytic'' tools for tackling this important communication-theoretic problem.\\
The paper is organized as follows: Section \ref{b} provides some basic definitions and is followed by Section \ref{sec-bounded-support} which contains the main result of this paper. As mentioned above, in Section \ref{sec-discussion} we give an elementary example that shows that the application of the identity theorem in higher dimensions is, in general, not admissible if we want to understand the properties of capacity maximizers of Rayleigh fading channels. \\
\emph{Notation.} Throughout the paper we will denote the set of complex $N$-by-$1$ matrices by $\mm(N\times 1,\cc) $ and will freely identify this set with $\cc^{N}$. $\ln$ stands for the logarithm to the base $e$. Capital letters $X,Y,H$ are reserved for random variables.
\section{Rayleigh fading channel}\label{b}
We consider a Rayleigh fading channel with the coherence time $T=1$ which is described by
\begin{equation}
Y_{m}=\sum^N_{n=1} H_{mn} X_{n} + Z_{m} 
\end{equation}
with coefficient matrices $Y,Z \in \mm(M\times 1,\cc)$,\\$X \in \mm(N\times 1,\cc)$ and $H \in \mm(M\times N,\cc)$, where the the channel H is assumed to be complex circularly symmetric Gaussian with zero mean and with covariance matrix $\Sigma$ and the additive noise coefficients $Z_m$ are assumed to be i.i.d. complex circularly symmetric Gaussian with $\cc\mathcal{N}(0,\sigma^2_Z)$.
Let $\mathcal{P}(X)$ be the set of probability measures on \\$(\mm(N\times 1,\cc), \Sigma_{Borel}(\mm(N\times 1,\cc)))$. Then the set 
\begin{equation}
\mu_{g,a}(X)=\{\mu \in \mathcal{P}|\int (g(x)-a)d\mu(x) \leq 0\}
\end{equation}
with the average power constraint of the transmitted signal 
\begin{equation}
\int (g(x)-a)d\mu(x)=\int \frac {1}{N} \sum^N_{n=1} |x_{n}|^2 d\mu(x)- a \leq 0
\end{equation}
is weak* compact as it was shown in \cite{fozunbal-a} and \cite{csiszar}.\\
If $\mathcal{P}(Y)$ is the set of conditional probability measures on \\$(\mm(M\times 1,\cc), \Sigma_{Borel}(\mm(M\times 1,\cc)))$ we can determine the channel by a set $\{W(\cdot |x) \in \mathcal{P}(Y)| \, x \in \mm(N\times 1,\cc)\}$, where $W(\cdot |x)$ is absolutely continuous with respect to Borel-Lebesgue measure. For the Rayleigh fading channel the conditional probability density of the received signals $y$ conditioned on the input symbol $x$ is given by
\begin{equation}\label{channel-def}
p(y|x)=\frac {e^{-\textrm{tr}\left[ (\sigma^2_Z \idn_{M} +(\idn_M\otimes x^H)\Sigma(\idn_M\otimes x))^{-1} yy^H \right ]}}{\pi^{M}\textrm{det} (\sigma^2_Z\idn_{M} + (\idn_M\otimes x^H)\Sigma (\idn_M\otimes x ))}
\end{equation}
with covariance matrix $\Sigma$ of $H$
\begin{equation}
\Sigma = \mathbf{E}(H\otimes H^*).
\end{equation} 
Let $\mu \in \mu_{g,a}(X)$ be a probability measure and define
\begin{equation}\label{int-measure}
f_{\mu}(y):=\int p(y|x) \mu(dx).  
\end{equation}
Then the mutual information of the channel with no CSI at the receiver is given by
\begin{equation}
I(\mu;W)=\int p(y|x) \log \frac{p(y|x)}{f_{\mu}(y)}dyd\mu(x).
\end{equation}
The mutual Information is a weak* continuous functional on the weak* compact and convex set $\mu_{g,a}(X)$ (see \cite{fozunbal-b}). Thus the functional $I(\mu;W)$ achieves its maximum on $\mu_{g,a}(X)$ by the following 
\begin{theorem}[Cf. \cite{abou-faycal}]
Let $f$ be a weak* continuous real-valued functional on a weak* compact subset $S$ of $X^*$. Then $f$ is bounded on $S$ and achieves its maximum on $S$.
\end{theorem}
The mutual information is strictly concave functional on $\mu_{g,a}(X)$ up to \emph{equivalence} of measures. Hereby, two measures $\mu,\nu \in \mu_{g,a}(X)$ are called equivalent if ${f_{\mu}(y)}={f_{\nu}(y)}$. So its maximum on $\mu_{g,a}(X)$ is achieved by a unique input distribution up to equivalence defined above \cite{fozunbal-b}. Hence, with
\begin{equation}
C(a)=\sup_{\mu \in \mu_{g,a}(X)}I(\mu;W)
\end{equation}
there exists a measure $\mu_0 \in \mu_{g,a}(X)$ that achieves the capacity of the channel and is unique up to equivalence of measures. The aim of this paper is to show that subjected to an  average power constraint the capacity achieving distribution of the channel has an bounded support. 
\section{Bounded support of optimal input distribution}\label{sec-bounded-support}
The purpose of this section is to show that the support of the capacity achieving input measure for the channel given in (\ref{channel-def}), with coherence time $T=1$, is bounded.\\
For $r_1,r_2 \in \rr$ with $0\leq r_1 < r_2$ we set
\begin{equation}
B (r_1,r_2):=\{ x \in \mm (N\times 1,\cc) : r_1\leq\textrm{tr}(xx^H)\leq r_2\},
\end{equation}
with $\langle x,x \rangle := \textrm{tr} (xx^H)=\| x\|^2$.\\
\begin{lemma}\label{lemma1}
Let $r_1,r_2 \in \rr$ with $0\leq r_1 < r_2$ and $\mu(B(r_1,r_2))>0$ with $\mu \in \mu_{g,a}(X)$ be given. Then
\begin{multline}
\int p(y|x)\log f_{\mu}(y) dy \\
\geq \log \frac{\mu (B(r_1,r_2))}{\pi^{M}\Pi}-\frac{M(\sigma^2_Z + \lambda_{min}x^Hx)}{(\sigma^2_Z +\lambda_{max}r_1)}
\end{multline} 
with $\Pi:=\max_{x\in B(r_1,r_2)}\textrm{det}(\sigma^2_Z \idn_{M} + (\idn_M\otimes x^H)\Sigma (\idn_M\otimes x))$ and $\lambda_{min}>0$ and $\lambda_{max}>0$ are the minimum and maximum eigenvalues of the covariance matrix $\Sigma$.
\end{lemma}
\begin{proof}
By the defining relation (\ref{int-measure}) we have
\begin{equation}\label{ineq-shell}
f_{\mu}(y):=\int p(y|x) \mu(dx) \geq \int_{B(r_1,r_2)} p(y|x) \mu(dx)
\end{equation}
Next we define\\
\begin{equation}
\Pi:=\max_{x\in B(r_1,r_2)}\textrm{det} (\sigma^2_Z \idn_{M} + ( \idn_M\otimes x^H)\Sigma (\idn_M\otimes x)),
\end{equation}
whereas the maximum of the function is achieved on $B(r_1,r_2)$ because of the compactness of $B(r_1,r_2)$. Hence, \\for $x \in B(r_1,r_2)$ we obtain
\begin{equation}\label{ineq-max}
p(y|x) \geq \frac {e^{-\textrm{tr}\left[ (\sigma^2_Z\idn_{M} +(\idn_M\otimes x^H)\Sigma(\idn_M\otimes x))^{-1} yy^H \right ]}}{\pi^{M}\Pi}.
\end{equation}
For every $x \in \mm(M\times 1,\cc)$ we have
\begin{multline}\label{ineq}
(\sigma^2_Z+\lambda_{min}x^Hx) \idn_M \\
\leq (\sigma^2_Z \idn_{M} +(\idn_M\otimes x^H)\Sigma(\idn_M\otimes x)) \\
\leq (\sigma^2_Z+\lambda_{max}x^Hx) \idn_M
\end{multline}
where $\lambda_{min}>0$ and $\lambda_{max}>0$ are the minimum and maximum eigenvalues of the hermitian and strictly positive covariance matrix $\Sigma$. By the definition of $B(r_1,r_2)$ we have
\begin{equation}
r_1\leq \textrm{tr}(xx^H)=x^Hx=\|x\|^2 \quad (x\in B(r_1,r_2)).
\end{equation}
Hence, it follows that
\begin{equation}
\sigma^2_Z + \lambda_{min}x^Hx\geq \sigma^2_Z + \lambda_{min}r_1 .
\end{equation} 
For two operators $A,B \in \mm(N,\cc)$ with $A \leq B$ and a positive operator $R \in \mm(N,\cc)$ we have 
\begin{equation}
\textrm{tr}(AR) \leq \textrm{tr}(BR).\\
\end{equation}
Due to the fact that the operators in (\ref{ineq}) are hermitian and positive and the same holds for $yy^H$ and because the function $f(A)=-A^{-1}$ is operator monotone for all positive operators \cite{bhatia}, we have
\begin{multline}
\textrm{tr}\left[((\sigma^2_Z + \lambda_{min} r_1) \idn_M)^{-1}yy^H\right ] \geq \\
\textrm{tr}\left[((\sigma^2_Z + \lambda_{min}x^Hx) \idn_M)^{-1}yy^H\right] \\
\geq \textrm{tr}\left[(\sigma^2_Z\idn_{M} +(\idn_M\otimes x^H)\Sigma(\idn_M\otimes x ))^{-1}yy^H \right].
\end{multline}
With (\ref{ineq-max}) it follows that for $x\in B(r_1,r_2)$
\begin{equation}
p(y|x)\geq \frac{e^{-\textrm{tr}\left[((\sigma^2_Z + \lambda_{min} r_1) \idn_M )^{-1}yy^H\right ]}}{\pi^{M}\Pi}
\end{equation}
Inserting this into (\ref{ineq-shell}) yields
\begin{equation}
f_{\mu}(y) \geq \frac {\mu(B(r_1,r_2))}{\pi^{M}\Pi} e^{-\textrm{tr}\left[((\sigma^2_Z + \lambda_{min} r_1 ) \idn_M)^{-1}yy^H\right ]}.
\end{equation}
Therewith we get
\begin{flalign}
&\int p(y|x)\log f_{\mu}(y)dy \geq &\nonumber   \\
&\int p(y|x)\log \left[ \frac {\mu(B(r_1,r_2))}{\pi^{M}\Pi} e^{-\textrm{tr}\left[((\sigma^2_Z + \lambda_{min} r_1 ) \idn_M)^{-1}yy^H\right ]} \right ] dy &\nonumber \\
&=\log \textrm{A} - \int \textrm{tr}\left[((\sigma^2_Z + \lambda_{min} r_1 ) \idn_M)^{-1}yy^H\right ] p(y|x) dy & \nonumber \\
&=\log \textrm{A} - \int \frac{\| y \|^2}{(\sigma^2_Z + \lambda_{min} r_1 )} p(y|x) dy &\nonumber \\
&=\log \textrm{A} - \frac{\textrm{tr} (\sigma^2_Z\idn_{M} +(\idn_M\otimes x^H)\Sigma(\idn_M\otimes x ))} {(\sigma^2_Z + \lambda_{min} r_1)} & \nonumber \\
& \geq \log \textrm{A} - \frac{\textrm{tr} (( \sigma^2_Z+\lambda_{max} x^H x ) \idn_M)} {(\sigma^2_Z + \lambda_{min} r_1)} & \nonumber \\
& = \log \textrm{A} - \frac{M(\sigma^2_Z +\lambda_{max}\| x\|^2)}{\sigma^2_Z +\lambda_{min} r_1} &
\end{flalign}
\begin{flalign}
&\textrm{with} \quad A:=\frac {\mu(B(r_1,r_2))}{\pi^{M}\Pi}.&\nonumber
\end{flalign}
\end{proof}
Determining the capacity achieving input distribution subjected to average power constraint is a convex optimization problem. Necessary conditions for the optimal input distribution can be derived from the local \emph{Karush-Kuhn-Tucker} conditions. Together with the fact that the mutual information is a concave functional and the convexity of the constraint functional we obtain (see \cite{luenberger} and \cite{fozunbal-b}), that $\mu$ achieves capacity if and only if 
\begin{equation}\label{loc-KKT}
\gamma (\frac{1}{N}\|x\|^2 - a)+C(a)-\int p(y|x)\log \frac{p(y|x)}{f_{\mu}(y)}dy \geq 0 
\end{equation}
with equality if $x\in \textrm{supp}(\mu)$, where $\gamma=\gamma (a) \geq 0$ denotes the Lagrange multiplier and
\begin{equation}
\int \frac {1}{N} \sum_{n} |x_{n}|^2 d\mu(x)\le a \nonumber
\end{equation}
is the constraint under consideration.
It is fairly standard fact that
\begin{multline}
\int p(y|x) \log p(y|x)dy= \\
-\log \left [(\pi e)^M \det(\sigma^2_Z\idn_M +(\idn_M \otimes x^H) \Sigma (\idn_M \otimes x))\right ] \nonumber
\end{multline}
and (\ref{loc-KKT}) can be therefore rewritten as
\begin{flalign}\label{loc-KKT3}
&\gamma (\frac{1}{N}\|x\|^2 - a)+C(a)+ \log(\pi e)^M +  & \nonumber \\
&+\log\det(\sigma^2_Z\idn_M +(\idn_M \otimes x^H) \Sigma (\idn_M \otimes x)) + &\nonumber \\
&+\int p(y|x)\log f_{\mu}(y)dy \geq 0 &
\end{flalign}
with equality if $x\in \textrm{supp}(\mu)$. Let
\begin{flalign}\label{KKT}
& KKT(x):= \gamma (\frac{1}{N}\|x\|^2 - a)+C(a)+ \log(\pi e)^M + & \nonumber\\
& +\log\det(\sigma^2_Z \idn_M +(\idn_M \otimes x^H) \Sigma (\idn_M \otimes x)) + &\nonumber\\
& +\int p(y|x)\log f_{\mu}(y)dy&
\end{flalign}
Then (\ref{loc-KKT}) can be rephrased as $KKT(x) \geq 0  $ for $x \in \mm(N\times 1,\cc) $ and $ KKT(x)=0$ if $ x\in \textrm{supp}(\mu) $.\\
The following theorem gives a sufficient condition for the boundedness of the support of the capacity achieving measure in terms of the Lagrange multiplier $\gamma$.
\begin{lemma}\label{lemma-bounded-input}
Let $a\in \rr_{+}$ be given and let $\mu$ be a capacity achieving input measure subject to the average power constraint $a$ for the channel (\ref{channel-def}). 
Then $\gamma(a)>0$ implies that $\textrm{supp}(\mu)$ is bounded.
\end{lemma}
\begin{proof}
The proof is by contradiction. Suppose that $\gamma(a)=\gamma>0$ and that $\textrm{supp}(\mu)$ is not bounded. By our assumptions we can find $r_1,r_2\in \rr$ with the following properties:
\begin{eqnarray}
\mu (B(r_1,r_2))>0 \\
\gamma - \frac{MN\lambda_{max}}{\sigma^2_Z +\lambda_{min}r_1}>0.\label{fac_greater_zero}
\end{eqnarray}
Applying \emph{Lemma} \ref{lemma1} to the function $KKT(x)$ defined in (\ref{KKT}) we obtain the following inequality.
\begin{flalign}\label{KKTb}
& KKT(x) \geq \gamma (\frac{1}{N}\|x\|^2 - a)+C(a)+ \log(\pi e)^M +   & \nonumber \\
& +\log\det(\sigma^2_Z\idn_M +(\idn_M \otimes x^H) \Sigma (\idn_M \otimes x))+ & \nonumber \\ 
& +\log A -\frac{M(\sigma^2_Z +\lambda_{max}\| x\|^2)}{\sigma^2_Z+\lambda_{min} r_1} & \nonumber \\
& = \|x\|^2 (\frac{\gamma}{N}- \frac{M\lambda_{max}}{\sigma^2_Z +\lambda_{min}r_1})-\gamma a+ C(a)+\log(\pi e)^M + & \nonumber \\
& +\log\det(\sigma^2_Z \idn_M +(\idn_M \otimes x^H) \Sigma (\idn_M \otimes x))+&\nonumber\\
& +\log A - \frac{M\sigma^2_Z}{\sigma^2_Z+\lambda_{min} r_1} &
\end{flalign}
Combining the Karush-Kuhn-Tucker conditions and (\ref{KKTb}) we obtain that for any $x\in \textrm{supp}(\mu)$
\begin{flalign}
& 0=KKT(x) \geq &\nonumber\\ 
& \|x\|^2 (\frac{\gamma}{N}- \frac{M\lambda_{max}}{\sigma^2_Z +\lambda_{min}r_1})-\gamma a+ C(a)+\log(\pi e)^M +& \nonumber \\
& +\log\det(\sigma^2_Z\idn_M +(\idn_M \otimes x^H) \Sigma (\idn_M \otimes x))+& \nonumber\\
& +\log A -\frac{M\sigma^2_Z}{\sigma^2_Z+\lambda_{min} r_1} &
\end{flalign}
But this last inequality with our assumption that $\textrm{supp}(\mu)$ is not bounded, (\ref{fac_greater_zero}), and the fact that 
\begin{flalign}
&\|x\|^2 (\frac{1}{N}\gamma- \frac{M\lambda_{max}}{\sigma^2_Z+\lambda_{min}r_1})\to \infty \quad \textrm{as} \quad x\to \infty & \nonumber \\
&\textrm{and} & \nonumber \\
& \log\det (\sigma^2_Z \idn_M + (\idn_M \otimes x^H) \Sigma (\idn_M \otimes x)) \to \infty \quad \textrm{as} \quad x\to \infty &\nonumber
\end{flalign}
implies that $0 \geq \infty$, which is the desired contradiction.
\end{proof}
In view of Lemma \ref{lemma-bounded-input} our remaining goal is to show that $\gamma(a)>0$ for each $a\in\rr_{+}$. For example in \cite{abou-faycal} Abou-Faycal, Trott, and Shamai showed this in the scalar case. Our proof of the corresponding result in MIMO case below is strongly motivated by their approach via Fano's inequality. 
\begin{lemma}\label{gamma>0}
For the channel given in (\ref{channel-def}) we have $\gamma(a)>0$ for each $a\in\rr_{+}$.
\end{lemma}
\begin{proof}
As mentioned above the proof is an extension of the argument given in \cite{abou-faycal}. The capacity functional $C(\cdot)$ is a non-decreasing and concave function of the argument $a\in \rr_{+}$. It was observed in \cite{abou-faycal} using \emph{global} Karush-Kuhn-Tucker conditions that $\gamma(a)$ is the slope of the tangent line to $C(\cdot)$ at $a$ (cf. \cite{abou-faycal}, Section III.B and Appendix II.A). Thus, since $C(a)$ is non-decreasing and concave, it can be shown that $\gamma(a)=0$ implies $C(a')=C(a)$ for all $a'\ge a$\footnote{This implication is not obvious since $C(\cdot)$ need not be differentiable. However, $C(\cdot)$ is differentiable a.e. due to the monotonicity and concavity. The proof that $C(a)=C(a')$ for all $a'\ge a$ follows a standard line of reasoning from the real analysis and is skipped due to the space limitation. The full argument will be given elsewhere.}. Consequently, we can rule out the possibility that $\gamma(a)=0$ by showing the existence of a sequence of input measures such that the corresponding sequence of mutual informations approaches $\infty$.\\
We will be done if there is $\lambda>0$ such that for each $n\in \mathbb{N}$ we can find distinct $x_1=x_1(n),\ldots,x_n= x_n(n)\in \cc^{N}$ and disjoint measurable sets $B_1=B_1(n),\ldots,B_n= B_n(n)\subset \cc^{M}$ such that
\[ \int_{B_i}p(y|x_i)dy\ge \lambda \]
for all $i=1,\ldots, n$. Because a simple application of Fano's inequality with block length $1$ shows then that for the input measures $\mu_n:=\frac{1}{n}\sum_{i=1}^{n}\delta_{x_i}$ ($\delta_{x_i}$ is the point measure concentrated on $x_i$) we have
\[I(\mu_n,W)\ge \lambda\log n -1.  \]
Now we define
\[ \lambda:=\frac{1}{2} \frac{\omega_{2M}\lambda_{min}}{2\pi^M\lambda_{max} }e^{-\frac{\sigma_Z^2+\lambda_{min}}{\lambda_{min}} } >0 \]
where $\omega_{2M}$ denotes the surface area of the unit sphere in $\cc^{M}\simeq\rr^{2M}$ and $\lambda_{min}, \lambda_{max}$ are the smallest and the largest eigenvalues of $\Sigma$ and let $n\in\mathbb{N}$ be given.\\
We will now present the construction of the vectors $x_1=x_{1}(n),\ldots ,x_n=x_n(n)\in \cc^{N}$ and the decoding sets $B_1=B_1(n),\ldots, B_n=B_n(n)$. Let $x\in \cc^{N}$ with $||x||=1$ be fixed and consider a large positive real number $K=K(n)\ge 1$ that will be specified later. Set $x_i:=K_i x$ for $i=1,\ldots, n$ where $K_i:=K^{2^{i}} $. \\
Let $\lambda_{min}$ denote the smallest eigenvalue of $\Sigma$. For $i=1,\ldots, n$ we set
\begin{equation}\label{gamma-0}
r_i=r_i(K):=\sqrt{\sigma_{Z}^2+\lambda_{min}}K_i  
\end{equation}
and $B_i:=D(r_i,r_{i+1})$ where
\[ D(r_i,r_{i+1})=\{ y\in \cc^{M}: r_i\le \textrm{tr}(yy^H)=\langle y, y\rangle < r_{i+1} \}.  \]
As shown in the proof of Lemma \ref{lemma1} we have
\begin{equation}\label{gamma-1}
p(y|x)\ge \frac{e^{-{\frac{\langle y,y\rangle}{\sigma_{Z}^2+\lambda_{\min}\|x\|^2}}}}{ \pi^{M}\det({\sigma_{Z}^2+\lambda_{max}\|x\|^2 \idn_M})} .  
\end{equation}
Using (\ref{gamma-1}) and transforming to spherical coordinates in $\cc^{M}\simeq \rr^{2M}$we obtain
\begin{eqnarray}\label{gamma-2}
 \int_{B_i}p(y|x_i)dy &\ge& \frac{\omega_{2M}}{\pi^{M}(\sigma_{Z}^2+\lambda_{max} K_{i}^{2})^{M}}\nonumber\\
&& \times \int_{r_i}^{r_{i+1}}e^{-a_i r^2} r^{2M-1}dr, 
\end{eqnarray}
where $\omega_{2M}$ denotes the surface area of the unit sphere in $\cc^{M}$ and $a_i=a_i(K):=\frac{1}{\sigma_{Z}^2 +\lambda_{min}K_i^2}$. After the substitution $t=a_i r^2$ in the integral on the RHS of the inequality (\ref{gamma-2}) we arrive at
\begin{eqnarray}\label{gamma-3}
  \int_{B_i}p(y|x_i)dy &\ge&  \frac{\omega_{2M}(\sigma_Z^2+\lambda_{min}K_i^2 )^{M}  }{2\pi^{M}(\sigma_{Z}^2+\lambda_{max} K_{i}^{2})^{M}}\nonumber\\
&&\times \int_{a_ir_i^2}^{a_ir_{i+1}^2}e^{-t}t^{M-1}dt.
\end{eqnarray}
In what follows we use the abbreviation
\begin{equation}\label{gamma-4}
  F(K_i):=\frac{\omega_{2M}(\sigma_Z^2+\lambda_{min}K_i^2 )^{M}  }{2\pi^{M}(\sigma_{Z}^2+\lambda_{max} K_{i}^{2})^{M}}.
\end{equation}
The defining relation (\ref{gamma-0}) and our assumption that $K\ge 1$ ensure that $a_ir_i^2\ge 1$. Using this and (\ref{gamma-3}) we are led to
\begin{eqnarray}\label{gamma-5}
   \int_{B_i}p(y|x_i)dy &\ge& F(K_i)\int_{a_ir_i^2}^{a_ir_{i+1}^2}e^{-t}t^{M-1}dt\nonumber\\
&\ge& F(K_i)\int_{a_ir_i^2}^{a_ir_{i+1}^2}e^{-t}dt\nonumber\\
&=& F(K_i)(e^{-a_ir_i^2}-e^{-a_ir_{i+1}^2}),
\end{eqnarray}
for all $i=1,\ldots,n$. Now, since $K_i=K^{2^i}$, $a_i=a_i(K):=\frac{1}{\sigma_{Z}^2 +\lambda_{min}K_i^2}$, and $r_i=r_i(K):=\sqrt{\sigma_{Z}^2+\lambda_{min}}K_i $ it is clear that
\[ a_ir_{i+1}^2\to \infty \textrm{ as }K\to \infty, \]
\[ a_ir_i^2= \frac{\sigma_Z^2+\lambda_{min}}{\lambda_{min}}, \textrm{ as }K\to\infty \]
and from (\ref{gamma-4}) we have
\[ F(K_i)\to \frac{\omega_{2M}\lambda_{min}}{2\pi^M\lambda_{max} }\textrm{ as }K\to\infty \]
for all $i=1,\ldots,n$. Thus if we choose our $K$ sufficiently large (\ref{gamma-5}) and these limit relations ensure that
\[ \int_{B_i}p(y|x_i)dy \ge\frac{1}{2} \frac{\omega_{2M}\lambda_{min}}{2\pi^M\lambda_{max} }e^{-\frac{\sigma_Z^2+\lambda_{min}}{\lambda_{min}} }=\lambda>0, \]
for all $i=1,\ldots,n$. Moreover it is clear that the sequence of second moments of the measures $\mu_n=\frac{1}{n}\sum_{i=1}^{n}\delta_{x_i}$ can be made arbitrarily large for large $K(n)$.
This concludes our proof by the remarks given at the beginning of the argument.
\end{proof}
Now, we can summarize our results obtained so far in the following fashion:
\begin{theorem}\label{theorem-bounded-input}
We consider the channel defined by (\ref{channel-def}). Then the support of the capacity achieving input measure is bounded.
\end{theorem}
\begin{proof}
Simply apply Lemma \ref{gamma>0} and Lemma \ref{lemma-bounded-input}.
\end{proof}
\section{Discussion}\label{sec-discussion}
With the embedding function $\xi :  \cc^N  \to \rr^{2N} \in \cc^{2N}$ with $z_i=\textrm{Re}(x_i)$ and $z_{i+1}=\textrm{Im}(x_i)$
and the transformed channel we get an extension of the function
\begin{flalign}
&KKT(x): \mm(N\times 1,\cc) \to \rr &\nonumber \\
&\textrm{to}&\nonumber \\
&KKT(z): \mm(2N\times 1,\cc) \to  \cc & \nonumber  \\
&\textrm{where} & \nonumber\\
&KKT(z):=\gamma (\frac{1}{N}z^Tz - a)+C(a)-\int \tilde{p}(\tilde{y}|z)\log \frac{\tilde{p}(\tilde{y}|z)}{f_{\mu}(\tilde{y})}d\tilde{y}. & 
\end{flalign}
$\tilde{p}$ and $\tilde{y} \in \mm(2M \times 1,\rr)$ are obtained by changing the channel matrix and the channel output according the transformation of the input under $\xi$ (in \cite{chan} p. 2081, \cite{fozunbal-a}). Moreover it is easily seen using Fubini's theorem from measure theory and Morera's theorem from the complex analysis in several variables (cf. \cite{vladimirov}) that this extension of the function $KKT$ is holomorphic.  But, unfortunately, it is \emph{not} true that the identity theorem (also known as the uniqueness theorem) holds for open sets in $\rr^{2N}$ as the following standard example shows:\\
\emph{Example.} We consider the simplest non-trivial case $\cc^2$. Let $\{e_1,e_2\}$ denote the standard basis of $\cc^2$ and let $f:\cc^2\to \cc$ be defined as
\[ f(z):=z^{T}e_2=z_1\cdot 0+ z_2\cdot 1=z_2 \]
where $^{T}$ denotes the transpose and $z_1,z_2$ are the coordinates of $z\in \cc^2$ with respect to the basis $\{e_1,e_2\}$. Clearly, $f$ is holomorphic and the set of zeros of $f$ is 
\[\mathcal{N}(f)=\{\cc\cdot e_1\}\simeq \rr^{2}.  \] 
In what follows we identify $\mathcal{N}(f)$ with $\rr^2$. $\rr^2$ is, by definition, open in the natural topology on $\rr^2$ (but it is \emph{not} open in the natural topology of $\cc^2$, it is a closed linear subspace of $\cc^2$), and the function $f$ is, apparently, not identically zero on $\cc^2$.\\
Note that this example with the identical arguments shows also that the conclusion of the identity theorem is not valid for open balls, say, in $\rr^{2}\subset\cc^2$. If $B\subset \rr^{2}\subset\cc^2$ is any open ball in $\rr^2$ then $f(z)=0$ for all $z\in B$ but, again, $f\neq 0$ on $\cc^2$. The reason is, as before, that an open ball in $\rr^2$ (with the natural topology of $\rr^2$) is not open in the topology of $\cc^2$.\\
This last example shows that the proof of Proposition 4.3 in \cite{fozunbal-a} is not correct, since it assumes the validity of the identity theorem in exactly this setting. It is this Proposition 4.3 in \cite{fozunbal-a} which would allow us to conclude that the support of the capacity achieving input measure contains no open sets (in $\cc^{N}\simeq\rr^{2N}$) provided we know that this support is bounded.\\
Actually, the authors of this paper are convinced that we need different mathematical techniques to tackle the problem of characterization of the optimal inputs for multiple antenna Rayleigh fading systems not relying on the identity theorem. One reason for this opinion is the fact that the characterization of sets for which the identity theorem holds (so called sets of uniqueness) in the setting of several complex variables is a long standing challenging open problem in complex analysis.
\section{Conclusions and future work}
We have shown that for a Rayleigh fading channel with coherence time $T=1$ the support of the capacity achieving input measure is bounded. Our method of proof does not allow to extend the results to the case $T>1$. In fact the techniques we have used have to be substantially sharpened and supplemented by additional new tools. Furthermore we have shown that the approach based on the application of the identity theorem from the complex analysis in several variables is not admissible. Therefore, it seems highly likely for us that the techniques needed should be ``real-analytic'' in spirit.


\section*{Acknowledgment}
This work is supported by the Deutsche Forschungsgemeinschaft DFG via project BO 1734/16-1 ''Entwurf von geometrisch-algebraischen und analytischen Methoden zur Optimierung von MIMO Kommunikationssystemen''.


\end{document}